# Extrinsic and intrinsic charge trapping at the graphene/ferroelectric interface

*M. Humed Yusuf, Bent Nielsen, M. Dawber and X. Du\**

Department of Physics and Astronomy, Stony Brook University, Stony Brook, New York 11794, USA

**ABSTRACT:** The interface between graphene and the ferroelectric superlattice $PbTiO_3/SrTiO_3$ (PTO/STO) is studied. Tuning the transition temperature through the PTO/STO volume fraction minimizes the adsorbates at the graphene-ferroelectric interface, allowing robust ferroelectric hysteresis to be demonstrated. "Intrinsic" charge traps from the ferroelectric surface defects can adversely affect the graphene channel hysteresis, and can be controlled by careful sample processing, enabling systematic study of the charge trapping mechanism.

Electric field gating of graphene using oxides with high κ values (e.g. $Al_2O_3$, $ZrO_2$) has been studied extensively[1,2], in the hope of enhancing performance over devices using more conventional dielectrics. These studies have been further extended to the use of the ferroelectric oxide $Pb(Zr_xTi_{1-x})O_3$ [3-5], which has a non-linear, hysteretic dielectric response to an electric field in addition to ultra-high κ (up to 3850). Carrier concentrations as high as $10^{13}$ cm$^{-2}$ [5] and mobility ~$10^4$ cm$^2$/Vs [3] have been reported for such systems at room temperature. Earlier works on graphene-ferroelectric systems demonstrated ferroelectric switching associated gating of graphene using ferroelectric polymer poly(vinylidene fluoride-trifluoroethylene) [P(VDF-TrFE)] in top-gated structures [6,7]. Further studies focused on combining graphene with high quality single crystal thin films, allowing atomically flat surface, large gate capacitance, and low switching voltage operations. In those devices however, an inverse resistance hysteresis (anti-hysteresis) of the graphene channel was typically observed [4,5,8]. The robust nature of this anti-hysteresis was attributed to water molecules [8] and large densities of $H^+$ and $OH^-$ ions [9], which screen the polarization between the graphene/PZT interface. More recently it was shown that the cleanliness of the graphene/PZT interface could be maintained by the transfer of graphene in deionized water instead of exposed to air [10]. In this work, the gating of graphene showed proper ferroelectric hysteresis, although the observation of the ferroelectric hysteresis strongly depended on ramping speed of the gate voltage. This indicates the presence of slow charge trapping processes, possibly associated with surface absorbates such as water molecules. Achieving a robust ferroelectric response, which is independent of ramping speed and range at room temperature, remains an open challenge for graphene-ferroelectric field effect transistors (GFeFETs).

This letter reports study of the graphene-ferroelectric interface using GFeFETs in which the ferroelectric material is an artificially layered PTO/STO ferroelectric superlattice. The advantage of using this kind of ferroelectric is that the ferroelectric transition temperature can be tuned by the PTO/STO volume fraction to obtain a clean graphene-ferroelectric interface with minimum surface absorbates. With proper growth conditions of the ferroelectric superlattices, ferroelectric resistance switching can be achieved reproducibly on the graphene channel. The switching is practically independent of the gate voltage ramping speed. With extrinsic contaminants and adsorbates largely removed, "intrinsic" charge traps from the ferroelectric substrate can still affect the behavior of the devices. The "intrinsic" traps are largely influenced by the growth parameters of the ferroelectric superlattices. Such "intrinsic" charge traps, while showing little impact on the capacitance-voltage (C-V) characteristics of the superlattice, strongly affect the resistance gating of the graphene channels. Compared to the adsorbate-associated traps, the surface defect-associated intrinsic traps show much faster time-response and can remain active even at cryogenic temperatures. Asymmetry of hole and electron trapping is observed. The charge trapping centers are attributed to physical defects on the surface of the superlattice.

An ideal substrate for a GFeFET should have high crystallinity, good ferroelectric properties and atomically smooth surfaces. As previously demonstrated [11], $PbTiO_3/SrTiO_3$ (PTO/STO) superlattices can be fabricated with a high degree of structural perfection and minimal dielectric leakage. Moreover, by varying the thickness of the PTO in the bi-layer, the transition temperature of a PTO/STO superlattice can be tuned. The transition temperature increases with increasing PTO volume fractions, and can be predicted by a simple electrostatic model and with Landau theory [11]. In this study, PTO volume fractions ranging from 0.5 to 0.86 were used ($T_C$ ranging from 127°C to 527°C) and so no significant effects arising from improper ferroelectricity [12] were expected. Since anti-hysteresis and ferroelectric switching on the graphene channel are competing processes, the above mentioned volume fractions ensure sufficient ferroelectric strength (between 0.34 Cm$^{-2}$ and 0.56 Cm$^{-2}$) from the superlattice to study the interplay between the two. For the present study, $n_p$ PTO / $n_s$ STO superlattices with varying thicknesses of PTO, $n_p$ unit cells thick, and STO, $n_s$ unit cells thick, were grown to a total thickness of ~100 nm on TiO$_2$ terminated SrTiO$_3$ substrates. Each superlattice was grown on top of a 20 nm SrRuO$_3$ (SRO) back electrode. Growth of the bottom electrodes and superlattices were through off-axis magnetron sputtering. Coherent growth, with a periodicity that results from the artificial layering in the structure, was evident in cross-sectional HRSTEM (Supporting Information Figure S1a) and x-ray diffraction measurements (Supporting Information Figure S1b). Atomic force microscopy (AFM) showed that the samples had good surface quality with RMS roughness of ~150 pm.

Following the growth of the ferroelectric superlattices, graphene was deposited at 250°C in ambient environment through mechanical exfoliation. Thin flakes of highly oriented pyrolytic graphite (HOPG) were placed onto the surface of the superlattice, sitting on a hot plate, and pressed down with a pressurized flow (around 30 psi) of high-purity nitrogen gas through a needle for 5s. The flakes were then blown away, leaving behind graphene that were optically identified. This process was repeated until a satisfactory number of graphene flakes were deposited on the surface of the superlattice (See Supporting Information). The temperature of 250°C was chosen in light of the previous results on successful removal of physisorbates from ferroelectric systems within the temperature range of 200°C - 350°C [14, 15], which enhanced the subsequent piezoresponse of these materials. This temperature is also instrumental in reducing the polarization of PTO/STO superlattices based on a previous work in characterizing these heterostructures at different temperatures [11]. For the superlattices in this work, an estimated 75% reduction in polarization is expected at 250°C; this further contributes to the reduction of surface adsorbates. Graphene can be treated thermally in ambient conditions at ~500°C without the emergence of defects [13], so no damage to the graphene should result from the deposition at 250°C. The graphene sheets used to make the devices were identified as being either single or few layer graphene (FLG) based on their optical contrast (Supporting Information, Figure S2a), and confirmed by contact mode AFM topography. GFeFETs were fabricated using standard e-beam lithography and metallization using palladium as contacts. The metal pads outside the graphene channel had $Al_2O_3$ insulation (150 nm thick) underneath in order to minimize current leakage through the ferroelectric thin film (inset of Figure S2b).

After the device fabrication, AFM imaging of the devices was carried out for studying the surface topography. Finally, the exposed graphene was protected from ambient adsorbates by a 150 nm thick coating of Poly(methyl methacrylate) (PMMA). PMMA is not expected to drastically influence the properties of graphene [16]. A schematic of the complete device structure is shown in Fig. 1a.

With graphene closely conforming to the surface of the ferroelectric superlattice, contact mode AFM topography on a finished device (Fig. 1b) provides a way to visualize the surface quality of the ferroelectric superlattice at the graphene-ferroelectric interface. The image on top of the graphene sheet is clear because the surface residues can be easily removed by the AFM tip [17]. These images reveal the clean interface between graphene and the superlattice. Devices with ideal interfacial topography, as exemplified by Figure 1b, will be discussed first. Here the superlattice beneath graphene shows unit-cell steps with atomically smooth surfaces, resulting from the miscut of the STO substrate.

These devices were characterized by measuring the two-terminal resistance as a function of the back-gate voltage. A relatively large contact resistance was consistently observed in all the devices with graphene deposited at 250$^o$C. This may be attributed to the greater amount of contaminants between graphene and the contact metal, mostly due to the strong electric field from the polarized ferroelectric substrate penetrating through the graphene. By contrast, devices made from graphene deposited at room temperature show much lower contact resistance; the contaminant layer between the substrate and graphene screens the electric field.

Figure 2a shows measurements performed on a graphene channel on a 6 PTO / 2 STO superlattice at 300K. A resistance hysteresis is observed in the graphene channel consistent with the ferroelectric domain switching of the superlattice, as the back gate voltage ($V_{BG}$) is swept from 4V to -4 V and from -4 V to 4 V (marked as red and black lines respectively). In a separate measurement on the same substrate (Figure 2b), using an isolated (85um X 85um) palladium top electrode in proximity of the graphene device, the capacitance of the ferroelectric layer was measured as a function of applied bias over a range of ± 4V. This indicates that a change in the ferroelectric coercive field of 8.6 MVm$^{-1}$ is sufficient to switch the polarization of the superlattice and confirms that the separation of the Dirac peaks is correlated with the switching of the polarization. Remarkably, the ferroelectric hysteresis behavior in these devices does not depend significantly on the ramping speed of the gate voltage (Figure 2c), indicating that there is no significant slow charge trapping process within the typical time scale of the measurements (from less than 1 min up to ~20 min for a completely ramped loop). The strength of this switching behavior increases with increasing coercive field of the superlattice (Inset of Figure

2c). It is noticed that in all our devices, there is a shift of the gating curves and C-V curves to the positive gate voltages, which is attributed to an as-grown down-polarization, which is commonly observed in sputtered PbTiO$_3$ thin films [18] and PbTiO$_3$ based superlattices [19].

Observation at low temperatures showed a continued increase in the Dirac peak separations with a steep increase for temperatures less than 55K (Figure 2d). This can be explained by the dynamics of the ferroelectric domain wall motion at these temperatures. It has been demonstrated previously in PZT systems [20, 21] that low temperatures restrict domain wall mobility, resulting in increasing coercive fields, $E_c$, needed for complete 180° domain reversal. Below 55K, a dramatic decrease in the 180° domain wall mobility makes it energetically costlier to switch the domains completely in this case.

The ferroelectric hysteresis on the graphene channel is reproducible with high-temperature graphene deposition and with proper choice of superlattice growth conditions; however, changing fabrication conditions can produce "non-ideal" devices that display anti-hysteresis characteristics. These "non-ideal" devices provide an opportunity for studying the impact of various extrinsic and intrinsic factors on the graphene-ferroelectric interface. To demonstrate superlattice defect-induced anti-hysteresis, the surface property of a 15 PTO / 3 STO superlattice was tuned by varying two parameters. The first approach was to artificially introduce defects on the superlattice top layer by holding the sample at 560°C for 2 hours, in growth atmosphere, after the growth was completed. The second was to vary the SRO (SrRuO$_3$) back-electrode growth temperature to produce SRO surfaces with different morphologies. Except for these two parameters, the rest of the fabrication process was kept unchanged. Figure 3 shows a comparison between the optimized growth condition and two other cases where the parameters mentioned above were changed. One can clearly observe that changing the growth conditions, as addressed above, caused a significant change in the surface topography of the graphene-ferroelectric interface, leading to single unit-cell (~0.4 nm) deep square-like pits. Figure 3(a) shows the surface of a superlattice that was grown on non-ideal/rougher SRO and was subsequently kept at the growth temperature for 2 hours after growth. Figure 3(c) shows a surface that resulted only by growing the superlattice on rougher SRO, without the additional step of keeping the superlattice at its growth temperature; the immediate effect was a decrease in the size and density of the pits. Figure 3(e) shows an ideal superlattice surface that was a result of growing the superlattice on SRO with atomically flat terraces, and the omission of the step of keeping the superlattice at its growth temperature. The density and size of the pits depend on both the quality of the SRO back electrode, and more severely, on the time for which the sample had been held at high temperature after growth. The important observation is the correlation of the surface quality of the ferroelectric superlattice and the gating behavior of the FET devices. With decreasing density/size of the pits, the devices switch from anti-hysteresis-like to ferroelectric-like, as demonstrated in Figures 3b, 3d & 3f. Such correlation was observed in all of the devices. This observation strongly suggests that surface defects play a vital role in determining the device characteristics of the GFeFETs.

Anti-hysteresis driven by the surface defects can be enhanced by increasing the range of $V_{BG}$, as is evident from anti-hysteresis of a 15 PTO / 3 STO sample at room temperature (Figure 4a). Screening of the polarization by traps between the graphene/ferroelectric interface is a result of the Fermi energy of the charge carriers, $E_F$, reaching the activation energy of the charge traps, $E_T$ [22]. This process becomes increasingly enhanced as more charges are fed into the different species of charge traps by increasing the maximum range of $V_{BG}$. The onset of charge trapping occurs at relatively small gating ranges (Figure 4a), after which the traps saturate and the strength of anti-hysteresis no longer increases. Moreover, the position of the Dirac peak, in the forward sweep, shows little ramping range dependence; this holds true for all the samples that underwent the high-temperature graphene deposition. On the other hand, the Dirac peak position in the reverse ramp sweep is strongly influenced by the ramping range in all of the devices. This behavior is indicative of an asymmetric electron and hole trapping mechanism in these devices. In comparison, for the devices with graphene deposited at ambient temperature, the characteristics are similar to what was previously reported on CVD graphene on PZT substrates [5]: the anti-hysteresis behavior persists for all the gate voltage ramping ranges, with electron and hole trapping being roughly symmetric (Supporting Information Figure S3). This is demonstrated by significant shifts of both the forward and backward ramped Dirac peaks, with $V_{BG}$ range, for such samples. By comparing the typical adsorbate-associated charge-trapping behaviors with the ferroelectric-dominated gating characteristics in high-temperature-deposited graphene samples, it is evident that high temperature deposition has reduced the effect of surface adsorbates.

Compared to what was previously reported in literature [10], which demonstrated devices changing from anti-hysteretic to ferroelectric hysteretic behavior by increasing the gate voltage ramping rate from 0.0187 V/sec to 1.0 V/sec, the charge-trapping process associated with the observed anti-hysteresis in this case is much faster. In these devices, changing the sweeping rate of the back-gate voltage within the conventional time-scales of sub-Hz, does not appear to influence the strength of the anti-hysteresis (Figure 4b), similar to the ferroelectric-hysteresis samples (Figure 2c). The charge trapping and the associated anti-hysteresis only appear to decrease significantly for gating frequencies higher than ~1KHz (~$10^4$V/sec) (Supporting Information, Figure S4a). Such a fast charge trapping process dominates the anti-hysteretic behavior of the R($V_{BG}$) dependence, while the slower extrinsic molecular process [4, 5, 8] typically associated with the presence of surface adsorbates/contaminants, barely play any role.

Contrary to the anti-hysteretic behavior observed in the graphene gating curves, capacitance measurements on palladium top-plate electrodes on the same samples always show proper ferroelectric-hysteresis behavior. This can be understood considering that capacitance measurements are only affected by the total charge induced on the top of the ferroelectric superlattice (including charges inside the graphene channel and charges that get trapped), as long as charge trapping is sufficiently fast compared to the measurement speed.

Figure 4c shows the example of a 9 PTO / 3 STO sample that exhibits crossing-over from anti-hysteresis to ferroelectric hysteresis around 26K. A similar "cross-over" process at low temperature has recently been seen by Rajapitamahuni et al [23], in graphene field effect devices based on (Ba,Sr)TiO$_3$ thin films. Temperature dependence of the gate modulation of graphene resistance is shown more elaborately in Figure 4d, keeping the back-gate sweeping range constant between ±8V. Here, the 6 PTO / 2 STO sample in Figure 1d is compared with two 9 PTO / 3 STO samples of varying qualities. All the samples have been subjected to high temperature deposition and the same fabrication techniques. It is remarkable that, within a sweeping range of ± 8V, the forward swept Dirac peak positions (indicated by black in Figure 4d) follow similar temperature dependence with similar quantitative values. However, the backward swept peak positions (indicated by red in Figure 4d) are strongly affected by the sample quality. This phenomenon again gives credence to an asymmetric electron and hole trapping mechanism.

The results obtained from a wide variety of graphene-ferroelectric superlattice FET devices fabricated with different parameters suggest: 1) The gate-dependence of the GFeFETs is determined by the competition between ferroelectric switching and charge trapping. 2) The surface adsorbates-associated charge trapping, which typically shows a slow time response as investigated previously, is largely reduced in these devices. 3) The charge trapping observed in anti-hysteresis devices with high-temperature graphene deposition is related to the surface defects, in particular, the square-shaped pits. 4) The trapping of electrons and holes is asymmetric.

The charge trapping process can be explained by the gating curves at various gate voltage ramping ranges, as schematically presented in Figure 5a. With increasing positive gate voltage, the induced electrons become partially trapped almost instantaneously (within much less than 100ms). The trapped electron density increases with increasingly positive gate voltages, as the electrons fill more and more states. Above a certain gate voltage, all the electrons traps are filled and no more electron trapping occurs. As the gate voltage is ramped down, on the other hand, the trapped electrons do not get de-trapped until a much lower $V_{BG}$ is reached, resulting in anti-hyteresis. (For more information on the de-trapping process, refer to Supporting Information Figure S4b)

It is also noticed that hole trapping has much weaker impact on the gating curves, as indicated by the weak ramping range dependence of the forward ramped Dirac peak. A possible explanation is that since the ferroelectric superlattice used in this work is down-polarized as fabricated, hole trapping happens constantly and therefore the hole traps are largely saturated by the time of measurements.

In Figure 5, the competition between charge trapping and ferroelectric switching is simulated. Assuming the charge traps are at a close vicinity to the channel, and the total gate-induced charge density $n_T$ goes to either the channel ($n_{ch}$) or the charge traps ($n_{tr}$):

$$\delta n_{tr} + \delta n_{ch} = \varepsilon \frac{\delta V_g}{ed} + \frac{\delta P(V_g)}{e} \quad (1)$$

Here $P(V_g)$ is the polarization of the ferroelectric gate dielectric. For $\frac{dV_g}{dt} > 0$, $P(V_g) = P_s \tanh(s(V_g - V_c))$ and for $\frac{dV_g}{dt} < 0$, $P(V_g) = P_s \tanh(s(V_g + V_c))$, where $P_s$ is the spontaneous polarization of the superlattice, taken to be 0.3 Cm$^{-2}$, $V_C$ is the coercive voltage taken to be 0.6V and $s$ is a fitting parameter that characterizes the sharpness of the ferroelectric switching. The dielectric constant, $\varepsilon$, is taken to be 250, as determined in our superlattices; $e$ is the electronic charge and $d$ is the thickness (100nm) of the superlattice thin film.

As the gate voltage is varied, the driving force for charge trapping is determined by the local electric field, which is in turn proportional to $n_{ch}$. Numerical simulations are carried out based on equation (1). Because charge trapping is much faster compared to the gate voltage ramping speed the details of its time dependence can be neglected. Therefore, $\delta n_{tr} = \beta(n_{ch}) \frac{dn}{dV_g} \delta V_g$ is assumed. Here, $\beta(n_{ch})$ describes the fraction of the induced charges $n$ that got trapped. The resistivity of the graphene channel is approximated as $R \sim \frac{1}{\sqrt{n_0^2 + n_{ch}^2}}$, where $n_0$ corresponds to a saturation carrier density due to potential fluctuation that prevents the resistivity from diverging numerically at the Dirac point.

The simple model above captures most of the experimental observations and provides a qualitative understanding of the impact of various factors on the Dirac point peak positions. Figure 5 shows examples on the key results. Assuming only electron traps play a role in the gating dynamics, it is found that the forward ramping peak position depends only weakly on the ramping range, while the reverse ramping peak position strongly depends on the ramping range (Figure 5b), consistent with what was observed in our measurements (Figure 4a.). At low temperatures (Figure 5c), the observed temperature dependence of the peak positions can be attributed to the temperature dependence of the coercive field, which increases with decreasing temperature.

Previous studies on the surfaces of $PbTiO_3$ and $Pb(Zr,Ti)O_3$ thin films can give some insight in to the nature of the surface charge traps. For example, Kim et al. [24] found a high density of negative trap states on the surface of a $Pb(Zr,Ti)O_3$ film using scanning Kelvin probe microscopy. In that paper, no images of surface topography were provided and hence it was not claimed that different regions of the film had different trap densities. Here, on the other hand, the density of traps correlates with the number and size of the pits on the surface, so it is reasonable to speculate that the pits are somehow different from the rest of the surface and host most of the negative charge traps responsible for the anti-hysteretic behavior. In surface x-ray diffraction studies of the $PbTiO_3$ surface by Munkholm et al. [25], it was found that the main decomposition process of the surface at high temperatures and low PbO pressure was an evolution from a well ordered 2x2 reconstruction towards a poorly ordered 1x6 reconstruction of the surface. The square pits that develop when the superlattices are left at high temperature post-deposition, when the chamber atmosphere contains no PbO, are most likely regions where the 1x6 reconstructed surface develops. It is suggested that these PbO deficient regions are responsible for the majority of the intrinsic traps that result in anti-hysteresis. The role of Pb and O vacancies in screening the ferroelectric polarization in PbO deficient $PbTiO_3$ thin films has been discussed by Selbach et al. [26].

In conclusion, it is shown that combining the tunability of ferroelectric superlattices and high temperature deposition of exfoliated graphene substantially reduces the level of interfacial adsorbates at the interfaces. The evidence of the reduction of interfacial adsorbates is the lack of slow charge trapping processes in the graphene-superlattice system. The surface morphology of the superlattices can be tuned through the growth conditions, which, in turn, determines the extent of the intrinsic charge trapping on the graphene channel and the nature of the channel resistance-voltage response of the device.

AUTHOR INFORMATION

Corresponding Author

*Email: * (X.D.) xu.du@stonybrook.edu

Author Contributions

M. Humed Yusuf was chiefly responsible for sample fabrication and measurement with assistance from B. Nielsen. M. Dawber and Xu Du were responsible for the design and direction of the project and prepared the manuscript with M. Humed Yusuf.

ACKNOWLEDGEMENTS

This work was supported by NSF (Grant : DMR 1105202). Part of this research was carried out at the Center for Functional Nanomaterials, Brookhaven National Laboratory, which is supported by the U.S. Department of Energy, Office of Basic Energy Sciences, under Contract No. DE-AC02-98CH10886. We thank Dong Su and Fernando Camino at CFN, for TEM work and assistance with the use of the cryogenic probe station. We also thank Niyati Desai for helping to proofread the manuscript.

FIGURES

**Figure 1.** (a) Schematic representation of a GFFET (graphene ferroelectric field effect transistor)**.** (b) Contact mode AFM (Atomic Force Microscopy) on the graphene device reveals 4 Angstrom high unit cell steps of the PTO ($PbTiO_3$) top layer underneath the graphene. The AFM topography confirms a clean interface between graphene and the ferroelectric surface with no signs of residues or bubble formation.

**Figure 2.** Ferroelectric switching on the graphene channel. Figures (a), (b) and (d) are 6/2 PTO/STO samples; figure (c) is a 15/3 PTO/STO sample. (a) $R$ vs $V_{BG}$ of a graphene sample on a 6/2 PTO/STO superlattice, measured in two-terminal configuration at room temperature. (b) Capacitance vs Applied bias measured across the superlattice. Samples, such as this one, show a direct capacitive coupling between the two materials and hence an excellent interface. (c) Peak positions from the $R$ vs $V_{BG}$ curves as function of ramping rate of a hysteretic graphene sample on a 15/3 PTO/STO superlattice. For sweeping speeds of sub-hertz frequencies, the hysteretic samples show no dependence of the peak positions with the speed of gating. Inset: Dirac peak positions of the same sample as a function of $V_{BG}$ sweeping range. (d) Low temperature characterization of a hysteretic graphene sample on a 6/2 PTO/STO superlattice, within a $V_{BG}$ of ±10V, shows subsequent enhancement of dirac peak separations with decreasing temperatures, revealing an increase in the coercive field of the superlattice. The gating curves are shifted for comparison at different temperatures.

**Figure 3**. The superlattice surface morphology can influence the graphene channel response. The left column shows contact mode AFM topography images, with the corresponding $R$ vs $V_{BG}$

curve for that sample shown directly to the right of each surface scan. The superlattices used are 15/3 PTO/STO. (a) Keeping a superlattice at high temperature (~ 560°C) in the growth atmosphere for 2 hours, right after growth, can create a surface with a large concentration of pits that are ~ 0.4 nm deep and ~ 90nm – 150 nm wide. (b) Accordingly, anti-hysteresis is observed in the graphene channel. (c) The superlattice surface quality was improved by immediately cooling the sample after the growth was completed. (d) Within a relatively high $V_{BG}$ of ±6.0V, the graphene channel response was predominantly hysteretic. (e) A near-perfect superlattice surface, which resulted from growing the superlattice on a SRO film with atomically flat terraces. (f) This gave rise to ferroelectric hysteresis on the graphene channel.

**Figure 4.** Anti-hysteresis on the graphene channel. Figures (a) and (b) are 15/3 PTO/STO samples, figure (c) is a 9/3 PTO/STO sample. (a) Peak positions as a function of the $V_{BG}$ ramping range of an anti-hysteretic graphene sample on a 15/3 PTO/STO superlattice at room temperature. Inset: Gating dependence on $V_{BG}$ sweeping range (from ±1.0V to ±4.0V) for the same sample; the curves are shifted for comparison at different gating ranges. (b) The peak positions show little variation below the conventional sub-hertz sweeping speeds, just like the hysteretic samples. This shows a fast trapping process. (c) Gate modulation of the graphene resistance on a 9/3 PTO/STO superlattice at various temperatures under a fixed gate ramping range of ± 8V; the gating curves are shifted for comparison at different temperatures. Crossing-over from anti-hysteresis to ferroelectric hysteresis happens as the temperature is lowered. (d) Dirac point peak position vs. temperature for forward and reverse gating for various samples. Remarkably, the forward gate ramping peak positions (black curves) were qualitatively and quantitatively similar for all the devices.

**Figure 5.** (a) A model of the charge trapping dynamics on the graphene channel. As an initial condition, for $V_{BG} = 0$, the superlattices have an intrinsic pinned-down self-polarization that populates the graphene with holes. Additionally, there is a specific density of hole traps at the interface; the density of hole traps remains the same throughout the cycle. Gradually, at a certain positive $V_{BG}$, switching happens and the superlattice is poled up; simultaneously, the electron traps become activated and deprive the graphene of some of the electrons. This brings forth the anti-hysteretic switching as observed in the backward ramped Dirac peak. As the gating is switched back to more negative values and the superlattice is poled down, the trapped electrons start to become de-trapped and gradually diminish in number, until there are none anymore. At this point, the graphene is populated with holes again and the whole process is reproducible with subsequent cycles. (b) Simulation showing graphene channel gating as a function of the gating range. The simulation assumes a fast-trapping process and confirms the experimental observation shown in Figure 4a. The curves are shifted for comparison at different gating ranges. (c) Simulation showing Dirac peak dynamics as a function of temperature; the gating curves are shifted for comparison at different temperatures and hence at different coercive fields. The simulation assumes that the crossing over is a result of an increase in the coercive field of the superlattice. This agrees with the experimental observation shown in Figure 4c.

# Figure 1

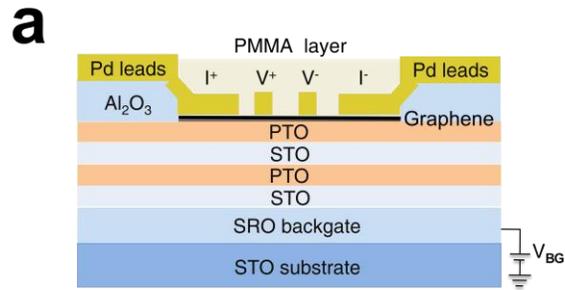

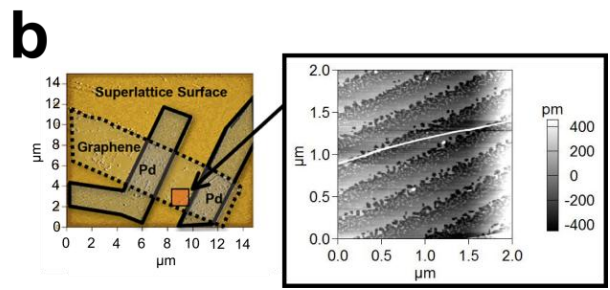

**Figure 2**

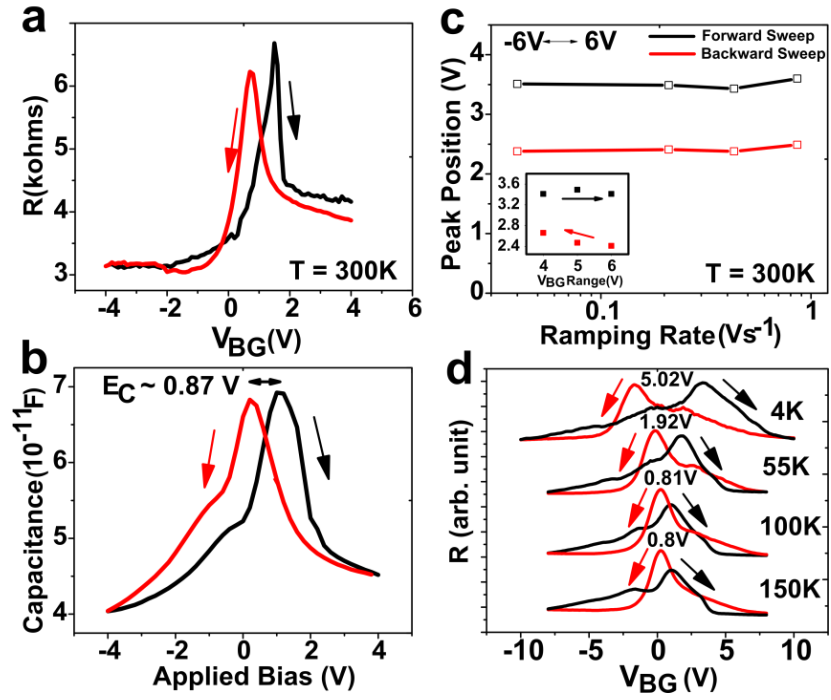

## Figure 3

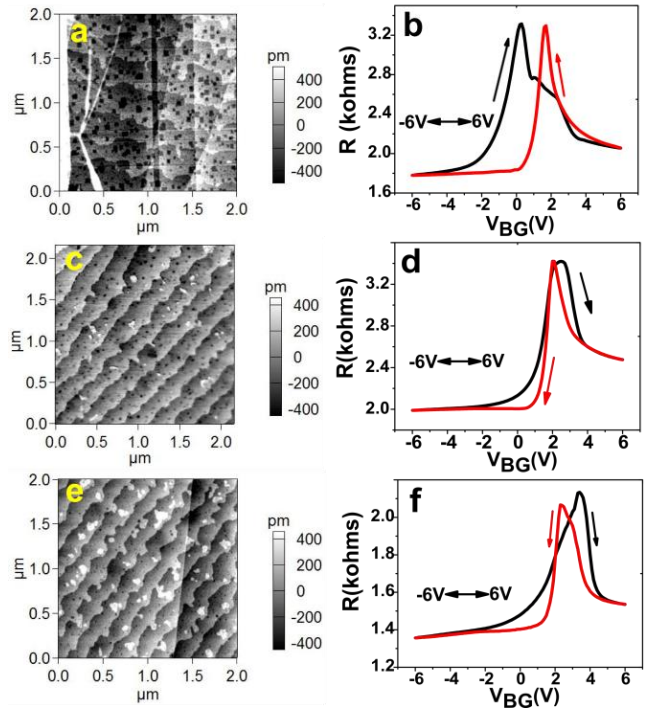

**Figure 4**

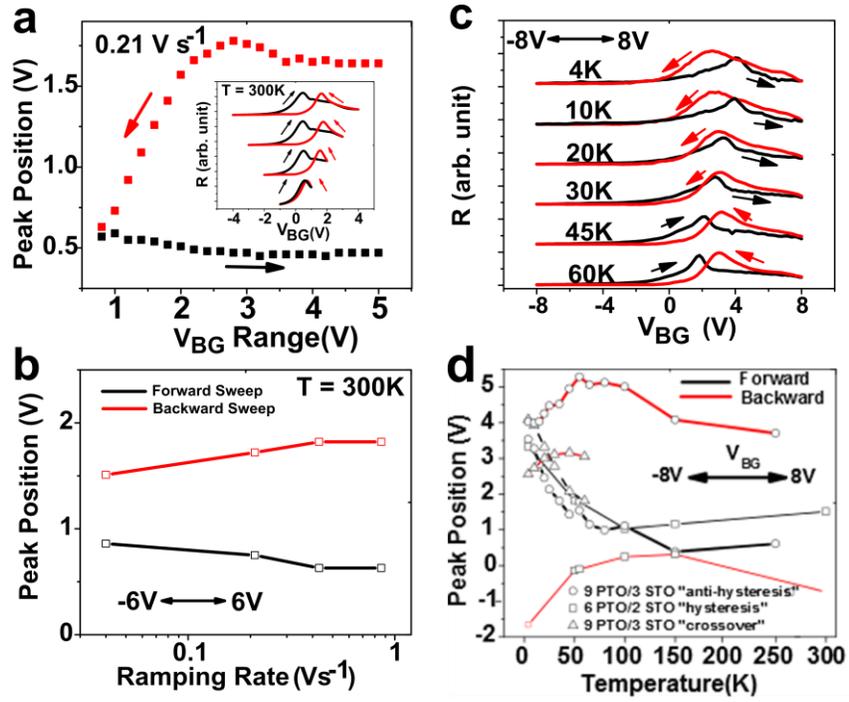

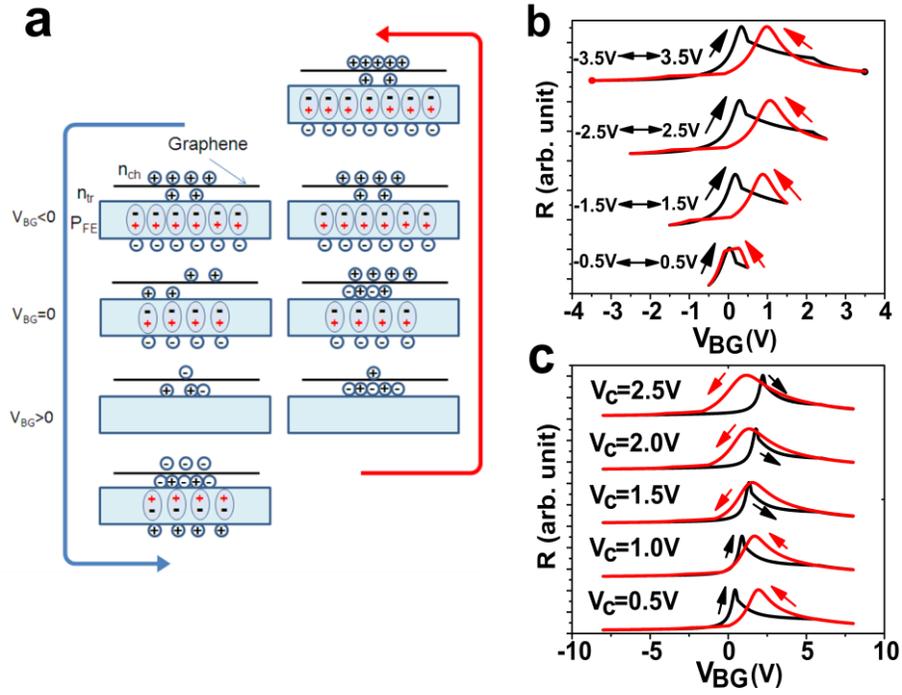